\documentclass[
    ,final            
  ]
  {aipproc}

\layoutstyle{6x9}

\usepackage{amsmath,amssymb}
\newcommand{\be}{\begin{equation}}
\newcommand{\ee}{\end{equation}}
\newcommand{\bea}{\begin{eqnarray}}
\newcommand{\eea}{\end{eqnarray}}
\newcommand{\bi}{\begin{itemize}}
\newcommand{\ei}{\end{itemize}}

\newcommand\hyn{\textit{\rm -}}

\begin{document}

\title{Study of charmonium-nucleon interaction\\ in lattice QCD}


\author{Taichi Kawanai and Shoichi Sasaki}{
  address={Department of Physics, The University of Tokyo, \\
Hongo 7-3-1, Tokyo 113-0033, Japan}
}

\begin{abstract}
We report preliminary results for charmonium-nucleon potential
$V_{c\bar{c} N}(r)$ from quenched lattice QCD,
which is calculated from the equal-time Bethe-Salpeter amplitude
through the effective Schr\"odinger equation.
Our simulations are performed at a lattice cutoff
of $1/a$=2.0 GeV in a spatial volume of $(3\;\text{fm})^3$ 
with the nonperturbatively $O(a)$ improved Wilson action
for the light quarks and a relativistic heavy quark action for the charm
quark. We have found that the potential $V_{c\bar{c} N}(r)$ 
 is weakly attractive at short distance 
and exponentially screened at long distance.
\end{abstract}

\maketitle


The heavy quarkonium state such as  the charmonium ($c\bar{c}$) state
does not share the same quark flavor with the nucleon ($N$).
This suggests that the heavy quarkonium-nucleon interaction is
mainly induced by the genuine QCD effect of 
multi-gluon exchange~\cite{{Brodsky:1997gh},{Luke:1992tm}}.
As an analog of the van der Waals force, 
two-gluon exchange contribution gives a weakly attractive, 
but long-ranged interaction between the heavy quarkonium state
and the nucleon. However, the validity of the calculation based on the 
perturbative theory is questionable for QCD where the nature of the strong coupling
appears in the long distance region.

The $c\bar{c}$-$N$ scattering at low energies has been studied 
from first principles of QCD. 
The $s$-wave $J/\psi$-$N$ scattering length 
is about 0.1 fm by using QCD sum rules~\cite{Hayashigaki:1998ey} and  $0.71\pm 0.48$ fm ($0.70\pm 0.66$ fm for $\eta_c$-$N$) 
by lattice QCD~\cite{Yokokawa:2006td}, while it is estimated as large as 0.25 fm 
from the gluonic van der Waals interaction~\cite{Brodsky:1997gh}. 
All studies suggest that the $c\bar{c}$-$N$ interaction is weakly attractive.
This indicates that the possibility of the formation of charmonium bound to nuclei
is enhanced.
In 1991, Brodsky {\it et al.} had argued that the $c\bar{c}$-nucleus ($A$) bound 
system may be realized for the mass number $A\ge 3$ if the attraction between 
the charmonium and the nucleon  is sufficiently strong~\cite{Brodsky:1989jd}. 
Therefore, precise information 
on the $c\bar{c}$-$N$ potential $V_{c\bar{c} N}(r)$ is indispensable for exploring 
nuclear-bound charmonium state like $\eta_c$-${}^{3}{\rm He}$ or
$J/\psi$-${}^{3}{\rm He}$ bound state in few body calculations~\cite{Belyaev:2006vn}. 

We recall a recent great success of the $N$-$N$ potential from lattice
QCD~\cite{Ishii:2006ec}.
In this new approach, the potential between hadrons can be calculated
from the equal-time Bethe-Salpeter (BS) amplitude through the effective
Schr\"odinger equation. Thus, the direct measurement of the
$c\bar{c}$-$N$ potential is now feasible by using lattice QCD. 
It should be very important to give a firm theoretical prediction about
nuclear-bound charmonium, which is possibly investigated by experiments 
at J-PARC and GSI. 


    The method utilized here to calculate the hadron-hadron potential
    in lattice QCD is based on the same idea originally applied for 
    the $N$-$N$ potential~\cite{{Ishii:2006ec},{Aoki:2009ji}}. 
    We first calculate the equal-time BS amplitude of two local 
    operators (hadrons $h_1$ and $h_2$)  
    separated by given spatial distances $r=|{\bf x}-{\bf y}|$ from the four-point correlator
    $
    G^{h_1\hyn h_2}({\bf r}, t_4, t_3; t_2, t_1)=\sum_{{\bf x}^{\prime}, {\bf y}^{\prime}}
    \langle {\cal O}^{h_1}({\bf x}, t_4){\cal O}^{h_2}({\bf y}, t_3)\left({\cal O}^{h_1}
    ({\bf x}^{\prime}, t_2){\cal O}^{h_2}({\bf y}^{\prime},t_1)    
    \right)^{\dagger}\rangle
    $, which becomes asymptotically proportional to $\phi_{h_1\hyn h_2}({\bf r}) e^{-E(t_3-t_1)}$
    for $|t_3-t_1|\gg 1$ with fixed $t_2$ and $t_4$, but keeping $|t_4-t_2|\gg 1$.
    Here $\phi_{h_1\hyn h_2}({\bf r})=\langle 0| {\cal O}^{h_1}
    ({\bf x}){\cal O}^{h_2}({\bf y})|h_1 h_2;E\rangle$ with the total energy $E$ for the ground 
    state of the two-particle $h_1\hyn h_2$ state corresponds to a part of the BS amplitude and 
    are called as the BS wave function~\cite{{Luscher:1990ux},{Aoki:2005uf}}. After an appropriate
projection with respect to discrete rotation of the cubic group, which is now ``rotational symmetry'' on the lattice, one can get the BS wave function projected in the $s$-wave. Once the BS wave
function $\phi_{h_1\hyn h_2}({\bf r})$ and the total energy $E$ are calculated in 
lattice simulations, the hadron-hadron potential can be obtained by
\be
V_{h_1\hyn h_2}({\bf r})=E+\frac{1}{2\mu}\frac{\nabla^2 \phi_{h_1\hyn h_2}({\bf r})}{\phi_{h_1\hyn h_2}({\bf r})}
\label{Eq.Pot}
\ee
where {\small$\mu$} is the reduced mass of the $h_1$-$h_2$ state and
 {\small $\nabla^2$} is defined by
the discrete Laplacian with nearest-neighbor points. More details of this method can be found in
Ref.~\cite{Aoki:2009ji}.

     In this study, we only consider the low energy $\eta_c \hyn N$ interaction, which 
     doesn't possess the spin dependent part. 
     We have performed quenched lattice QCD simulations 
     on two different lattice sizes, $L^3\times T=32^3\times 48$ and $16^3\times 48$,
     with the single plaquette gauge action at $\beta=6/g^2=6.0$, which corresponds
     to a lattice cutoff of $a^{-1} \approx 2.1$ GeV.
     Our main results are obtained from the data taken on the larger lattice ($La\approx$ 3.0 fm).
     A supplementary data with a smaller lattice size ($La\approx 1.5$ fm) are used for a test of 
     the finite size effect. The number of statistics is $O(600)$ for $L=32$ and $O(200)$ for
     $L=16$, respectively.
     
     We use non-perturbatively ${\cal O}(a)$ improved Wilson fermions 
     for the light quarks ($q$) 
     and a relativistic heavy quark (RHQ) action for the charm quark
     ($Q$)~\cite{Aoki:2001ra}.
     The RHQ action is a variant of the Fermilab approach~\cite{ElKhadra:1996mp}, 
     which can remove large discretization errors for heavy quarks.
     The hopping parameter is chosen to be
     $\kappa_q={0.1342,\ 0.1339,\ 0.1333}$, 
     which correspond to $M_\pi={0.64, 0.73, 0.87}$ GeV,
     and $\kappa_Q=0.1019$ which is reserved for the charm-quark mass 
     ($M_{\eta_c}=2.92$ GeV)~\cite{Kayaba:2006cg}. 
     Each hadron mass is obtained by fitting corresponding
     two-point correlation functions with a single exponential form.
     We calculate quark propagators with wall sources,
     which are located at $t_{\text{src}}=5$ for the light quarks 
     and at $t_{\text{src}}=4$ for the charm quark, with the Coulomb gauge fixing.
     The ground state dominance in the four point function
     is checked by the effective mass plot of the total
     energy of the $\eta_c$-$N$ system.

\begin{figure}
	\centering
	\includegraphics[width=.48\textwidth]{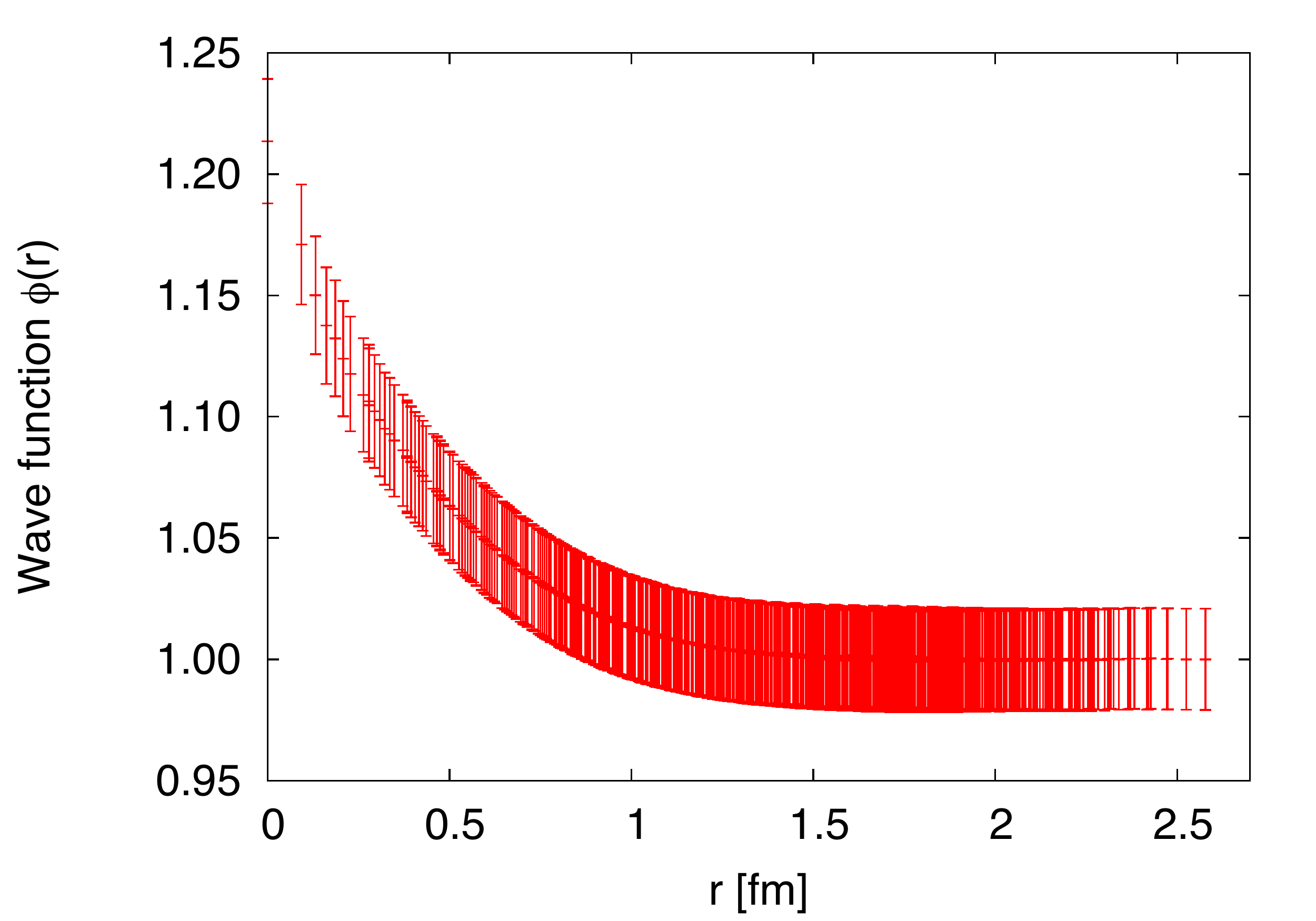}
	\includegraphics[width=.48\textwidth]{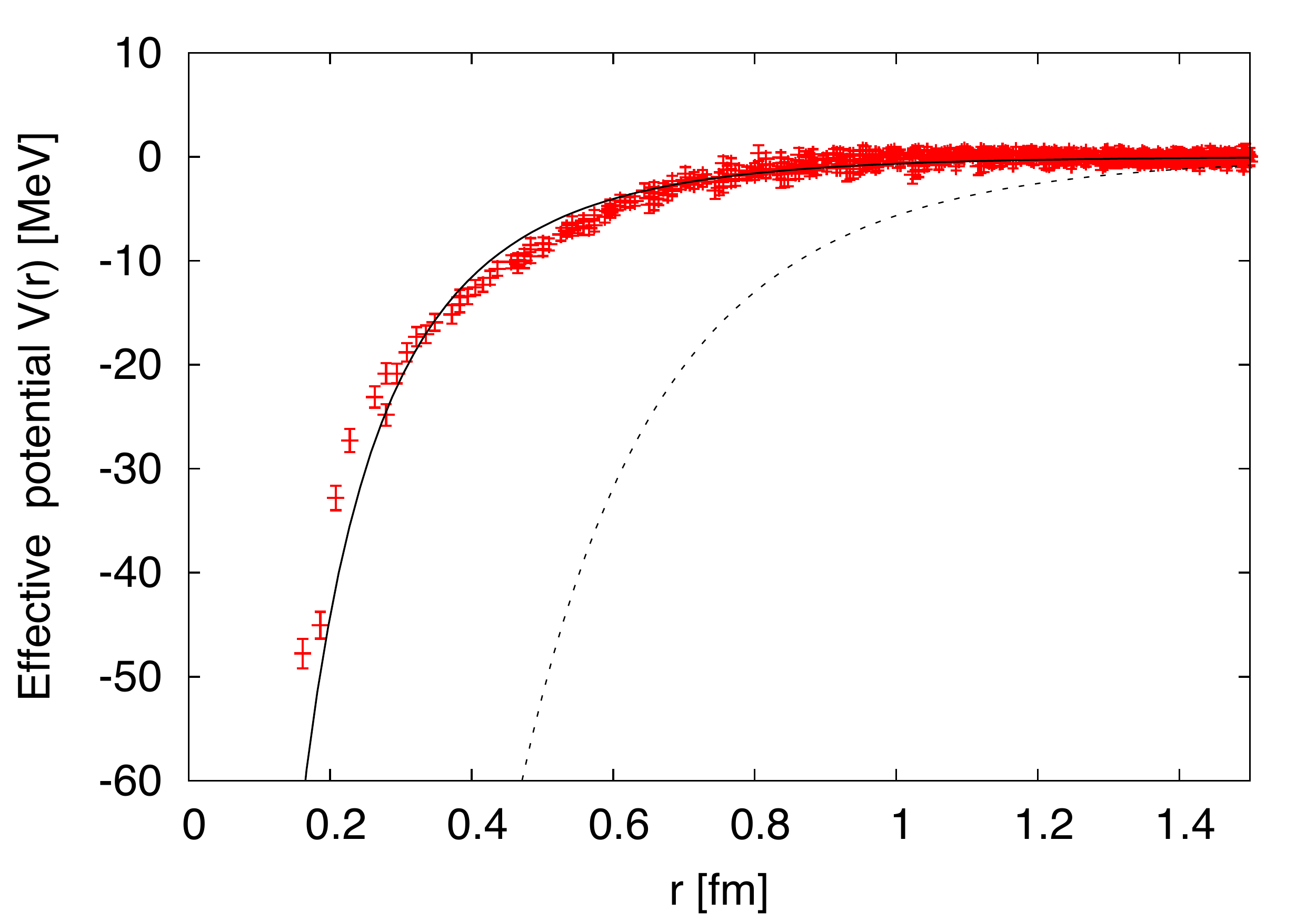}
	\caption{
	The wave function (left) and the effective central potential
	(right) in the $s$-wave $\eta_c$-$N$ system for $m_\pi= 0.64$
	GeV as a typical example. In the right panel, the solid 
	and dotted curves represent a fit result with the Yukawa form and
	the phenomenological potential adopted in Ref.~\cite{Brodsky:1989jd}, respectively.
	}
	\label{fig_results}
       \end{figure}

     The left panel of Fig.\ref{fig_results} shows a typical result
     of the projected BS wave function at the smallest quark mass,
     which is evaluated by a weighted average of data in the time-slice 
     range of $16 \leq t-t_{\text{src}}\leq 35$.
     The wave function is normalized to unity at a reference point ${\bf r}=(
     16,16,16 )$, which is supposed to be outside of the interaction region. 
     As shown in Fig.\ref{fig_results}, the wave function is enhanced 
     from unity near the origin so that the low-energy $\eta_c\hyn N$ 
     interaction is certainly attractive. This attractive interaction, however, 
     is not enough strong to form a bound state as is evident from this figure, 
     where the wave function is not localized, but extended at long distances.

      In the right panel of Fig.\ref{fig_results}, we show the effective central 
      $\eta_c\hyn N$ potential, which is evaluated by the wave function 
      through Eq.~(\ref{Eq.Pot}) with measured $E$ and $\mu$.
      As is expected, the $\eta_c\hyn N$ potential clearly exhibits 
      the entire attraction between the charmonium  and the nucleon 
      without any repulsion at either short or long distance. 
      It also can be observed that the interaction is exponentially screened 
      in the long distance region $r\gtrsim 1 \text{ fm}$.
      This is consistent with what we expected for the color van der Waals force
      in QCD theory, where the strong confining nature of the color electric field 
      must emerge~\cite{{Matsuyama:1978hf},{Feinberg:1979yw}}.
     
      
      In detail, the long-range screening of the color van der Waals force is confirmed  
      by the following analysis. We have tried to fit data with two types of fitting functions: 
      i) exponential type function as $-\exp(- r^m)/r^n$, which includes the 
      Yukawa form ($m=1$ and $n=1$), and ii) inverse power law  function as $-1/r^n$, 
      where $n$ and $m$ are not  restricted to be integers.
      The former case can easily accommodate a good fit with a small 
      $\chi^2$/ndf value, while in the latter case we cannot get any reasonable fit.
      For examples, functional forms $-\exp(- r)/r$ and  $-1/r^7$ give 
      $\chi^2/\mbox{ndf}\simeq 2.5$ and $34.3$ for fittings, respectively. 
       It is clear that
       the long range force induced by a normal ``van der Waals'' type potential based on 
       two-gluon exchange~\cite{Feinberg:1979yw} is non-perturbatively screened.
      
       If we adopt the Yukawa form $-\gamma e^{-\alpha r}/r$ to fit
       our data of $V_{c\bar{c}N}(r)$, we obtain $\gamma\sim 0.1$ and 
       $\alpha \sim 0.6$ GeV. These values should be compared with the phenomenological 
       $c\bar{c}$-$N$ potential adopted in Refs.~\cite{{Brodsky:1989jd}}, 
       where parameters ($\gamma=0.6$, $\alpha=0.6$ GeV) are barely fixed by a 
       Pomeron exchange model. The strength of the Yukawa potential
       $\gamma$ is six times smaller than the phenomenological one,  while the 
       Yukawa screening
       parameter $\alpha$ obtained from our data is comparable to the corresponding one.
       The observed $c\bar{c}\hyn N$ potential from lattice QCD is rather weak.

        \begin{figure}
	 \centering
	 \includegraphics[width=.48\textwidth]{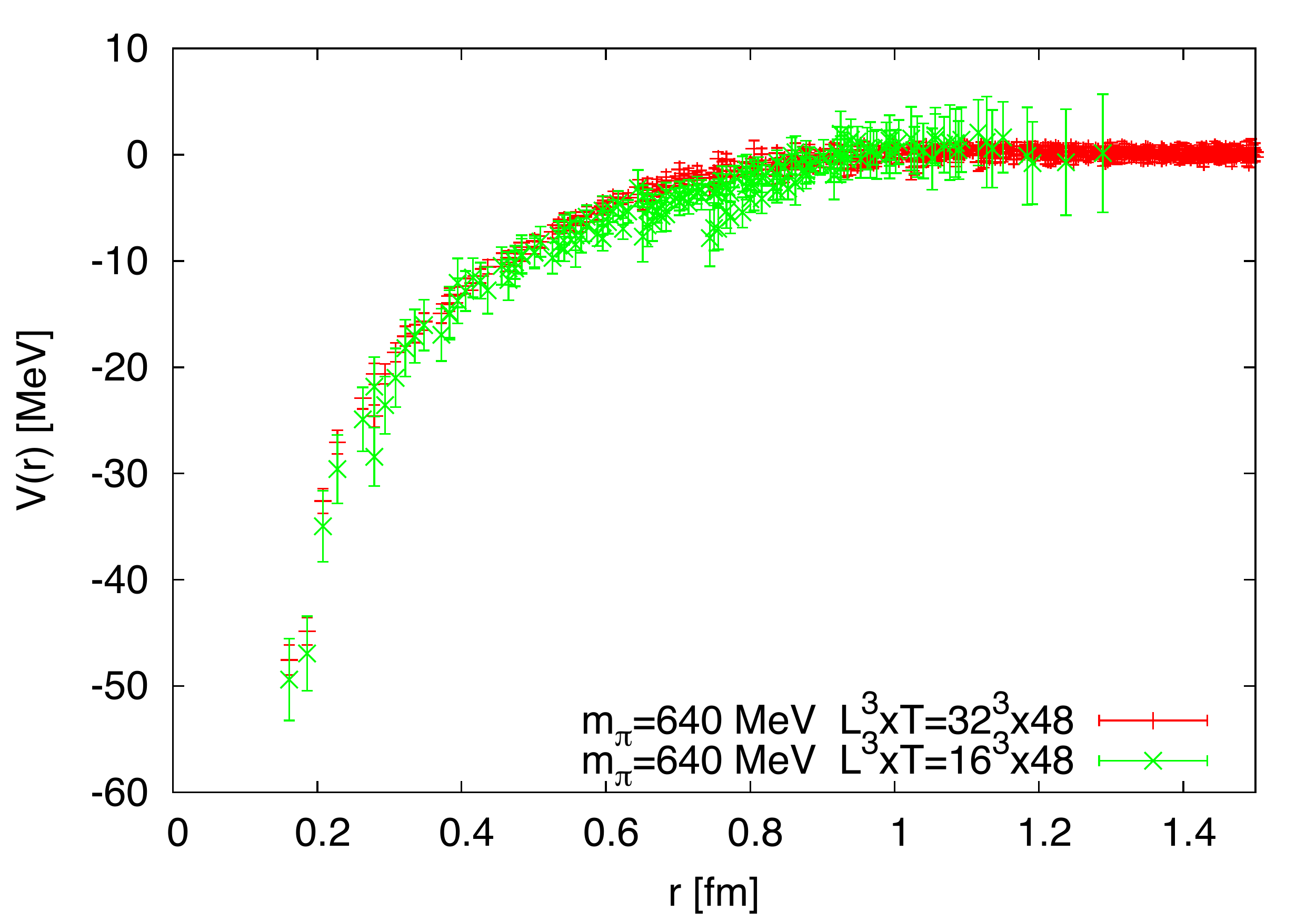} 
	 \includegraphics[width=.48\textwidth]{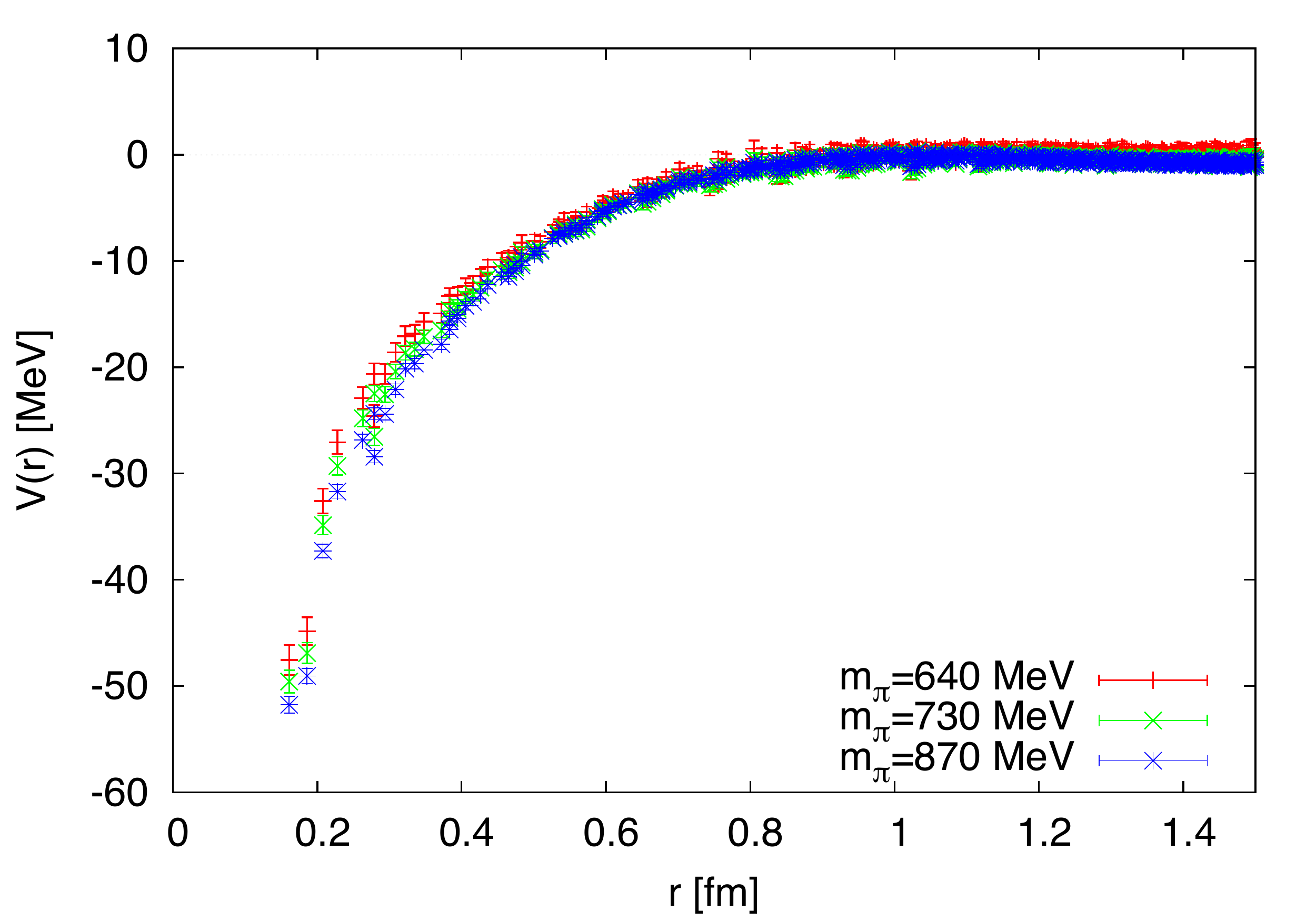}
	 \caption{The volume dependence (left) and the quark-mass dependence (right) on the $\eta_c\hyn N$ potential.}
	 \label{dependence}
	\end{figure}
      	
	We next show the finite size dependence and 
	the quark-mass dependence of the $\eta_c\hyn N$ potential in 
	Fig.~\ref{dependence}. 
	Firstly, as shown in the left panel of Fig.~\ref{dependence},
	there is no significant difference between
	potentials computed from lattices with two different spatial sizes ($La\approx 3.0$ and 1.5 fm).
	 This observation is simply because of the fact that the $\eta_c$-$N$ potential is 
	quickly screened to zero and turns out to be somehow short
	ranged. 
	In principle, the short range part of the potential, which is represented by the ultraviolet physics, 
	should be insensitive to the spatial extent associated with an infrared
	cutoff.
	As a result, it is assured that the larger lattice size is large enough to study 
	the $\eta_c$-$N$ system.		
	The appreciable quark-mass dependence is also not observed in 
	the right panel of Fig.~\ref{dependence}. This is expected from
	the fact that the $c\bar{c}\hyn N$ interaction is mainly 
	governed by multi-gluon exchange. However, it is worth mentioning that the ordinary 
	van der Waals interaction is sensitive to the size of the charge distribution. 
	Indeed, it is reminded that our simulations are performed
	in quenched approximation and at rather heavy quark masses. 
	This suggests that the $c\bar{c}$-$N$ potential 
	from  the dynamical simulations  would become more strongly attractive 
	in the vicinity of the physical point, where the size of the nucleon is much larger than
	at the simulated quark mass in this study. 
	    

We have studied the $c\bar{c}$-$N$ potential  $V_{c\bar{c}N}(r)$
from quenched lattice QCD, which is calculated from the equal-time 
BS amplitude through the effective Schr\"odinger equation.
It is found that potential  $V_{c\bar{c}N}(r)$ is weakly attractive at short distance 
and exponentially screened at long distance. In order to make a reliable
prediction about nuclear-bound charmonium, an important step in the future
is clearly an extension to dynamical lattice QCD simulation.
Such planning is now underway.

S.S. is supported by JSPS Grants (No. 19540265 and No. 21105504).
Numerical calculations reported here were carried out on
the PACS-CS supercomputer
at CCS, University of Tsukuba and also on the T2K supercomputer at ITC, University of Tokyo.



\bibliographystyle{aipproc}   


\begin{thebibliography}{9}

  
\bibitem{Brodsky:1997gh}
  S.~J.~Brodsky and G.~A.~Miller,
  Phys.\ Lett.\  B {\bf 412}, 125 (1997).

\bibitem{Luke:1992tm}
  M.~E.~Luke, A.~V.~Manohar and M.~J.~Savage,
  Phys.\ Lett.\  B {\bf 288}, 355 (1992).

\bibitem{Brodsky:1989jd}
  S.~J.~Brodsky, I.~A.~Schmidt and G.~F.~de Teramond,
  Phys.\ Rev.\ Lett.\  {\bf 64}, 1011 (1990).
  
\bibitem{Hayashigaki:1998ey}
  A.~Hayashigaki,
  Prog.\ Theor.\ Phys.\  {\bf 101}, 923 (1999).

\bibitem{Yokokawa:2006td}
  K.~Yokokawa, S.~Sasaki, T.~Hatsuda and A.~Hayashigaki,
  Phys.\ Rev.\  D {\bf 74}, 034504 (2006).

  
\bibitem{Belyaev:2006vn}
  V.~B.~Belyaev et al., 
  Nucl.\ Phys.\  A {\bf 780}, 100 (2006).
  
\bibitem{Ishii:2006ec}
  N.~Ishii, S.~Aoki and T.~Hatsuda,
  Phys.\ Rev.\ Lett.\  {\bf 99}, 022001 (2007).
  
\bibitem{Aoki:2009ji}
  S.~Aoki, T.~Hatsuda and N.~Ishii,
  Prog.\ Theor.\ Phys.\  {\bf 123} (2010) 89.
  
\bibitem{Luscher:1990ux}
  M.~L\"uscher,
  Nucl.\ Phys.\ B {\bf 354}, 531 (1991).

\bibitem{Aoki:2005uf}  
   S.~Aoki {\it et al.}  [CP-PACS Collaboration],
  Phys.\ Rev.\ D {\bf 71}, 094504 (2005).



\bibitem{ElKhadra:1996mp}
  A.~X.~El-Khadra, A.~S.~Kronfeld and P.~B.~Mackenzie,
  Phys.\ Rev.\  D {\bf 55}, 3933 (1997).

\bibitem{Aoki:2001ra}
  S.~Aoki, Y.~Kuramashi and S.~I.~Tominaga,
  Prog.\ Theor.\ Phys.\  {\bf 109}, 383 (2003).

    
\bibitem{Kayaba:2006cg}
  Y.~Kayaba {\it et al.}  [CP-PACS Collaboration],
  JHEP {\bf 0702}, 019 (2007).


\bibitem{Matsuyama:1978hf}
  S.~Matsuyama and H.~Miyazawa,
  Prog.\ Theor.\ Phys.\  {\bf 61}, 942 (1979).

\bibitem{Feinberg:1979yw}
  G.~Feinberg and J.~Sucher,
  Phys.\ Rev.\  D {\bf 20}, 1717 (1979).
  




\end{thebibliography}

\IfFileExists{\jobname.bbl}{}
 {\typeout{}
  \typeout{******************************************}
  \typeout{** Please run "bibtex \jobname" to optain}
  \typeout{** the bibliography and then re-run LaTeX}
  \typeout{** twice to fix the references!}
  \typeout{******************************************}
  \typeout{}
 }

\end{document}